# EXPLORING AMBIENT RADIO FREQUENCY EMISSIONS

**SUMMARY**: Radio astronomy observatories, such as the Dominion Radio Astrophysical Observatory (DRAO) in British Columbia, reside in 'radio quiet zones'. We explore how necessary such zones are by studying the radio frequency emission in our environment.

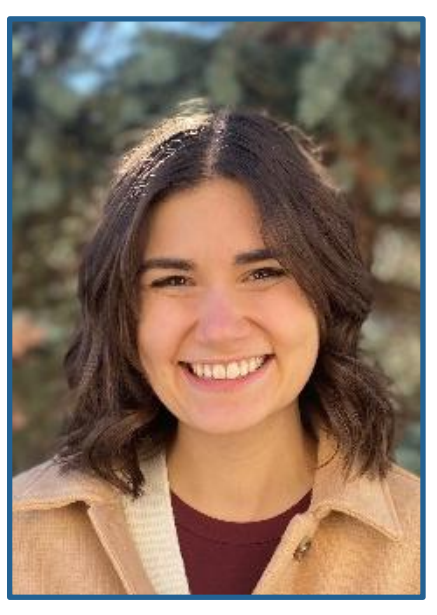

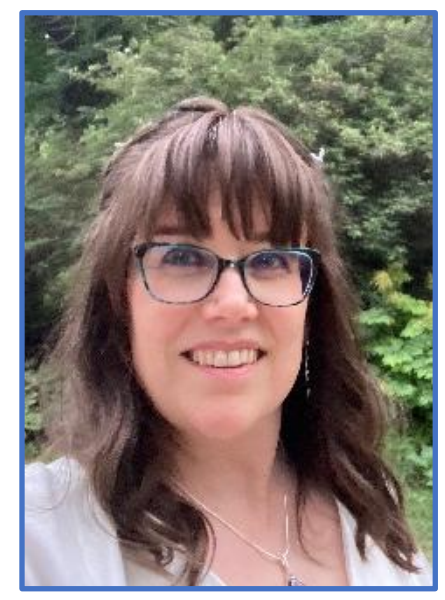

By **PAMELA FREEMAN** <pamela.freeman@ucalgary.ca> and **JO-ANNE BROWN** <jocat@ucalgary.ca>

Department of Physics and Astronomy, University of Calgary, Calgary AB T2N 1N4 Canada

Modern technology has led to radio frequency (RF) emission being virtually everywhere. The increasing reliance on services like cellular data, Wi-Fi, and Bluetooth, as well as the prevalence of older technologies such as microwaves, has led to inescapable daily RF exposure. The number of emitting devices is only growing as more and more uses of the wide range radio spectrum are discovered. The allocation of RF bandwidth for these different services is recommended by the International Telecommunication Union in the International Mobile Telecommunication (IMT) requirements and is determined at a national level where some frequency bands are reserved for astronomy and space physics research. The limited capacity of the radio band affects the expansion of current services and the development of new technology. Thus, there is always pressures to exploit frequencies currently protected for research.

The prevalence of RF emission has led to extensive investigation into its effects and potential uses. There are concerns around adverse health problems due to radiation exposure — although it is non-ionizing, RF radiation still may cause biological harm and government standards have been developed for exposure limits [1, and references therein]. Furthermore, as RF band crowding becomes more common, understanding possible unintentional interference becomes important. For example, band crowding can lead to compromised private communication, disrupted medical tests, or interrupted services [2].

RF interference (RFI) can be quite problematic in radio astronomy since it may lead to spurious observations and noise in the data. Protective measures for observatories include designated 'radio-quiet zones' and containing on-site instruments such as printers and hard drives in Faraday cages to limit artificial emission [3]. The Dominion Radio Astrophysical Observatory (DRAO) of the National Research Council of Canada (NRC; Figure 1) is in such a radio-quiet zone near Penticton, British Columbia. The DRAO contains the Synthesis Telescope and the 26-metre John A. Galt Telescope, which both operate at frequencies between 0.4 and 2 GHz. It also has a solar radio flux monitoring facility to provide space weather data for both research and commercial purposes. The site also hosts the Canadian Hydrogen Intensity Mapping Experiment, a novel, stationary radio telescope to map hydrogen in 3-D across most of the observable Universe. One of the main purposes of the DRAO, which it is internationally known for, is to design and develop radio instrumentation. DRAO monitors the RFI on site through a non-directional antenna but, as of 2024, there is no consistent recording outside of the CHIME band and a dedicated RFI monitoring system is still in progress [4].

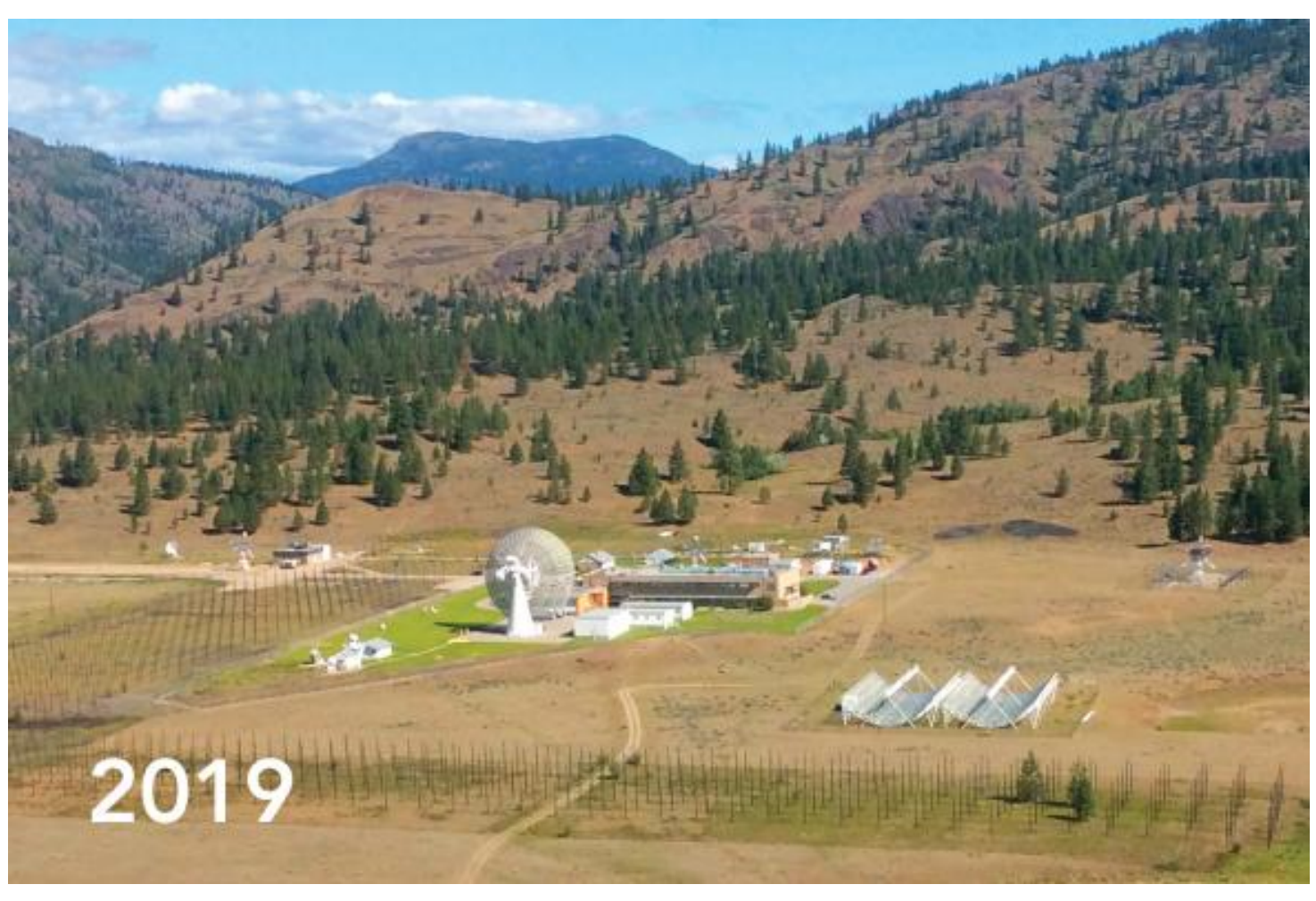

Figure 1. The Dominion Radio Astrophysical Observatory near Penticton, British Columbia. Image credit National Research Council.

Other observatories, such as the National Radio Astronomy Observatory (NRAO) in the United States, undertake the same precautions. A National Radio Quiet Zone in West Virginia was established around the site of the Green Bank Telescope, while the Jansky Very Large Array in New Mexico gains most of its protection from being geographically remote. The Atacama Large Millimeter/submillimeter Array, operated in part by the NRAO along with the European Southern Observatory and the National Astronomical Observatory of Japan, benefits both from a regulated protective zone and from its location at 5000 m altitude in the Atacama Desert of Chile. Despite any choice of site or of radio-quiet zones, these observatories still have to continuously monitor and mitigate human-made interference [e.g. 5,6].

Currently, most measures only include terrestrial-based sources of interference, however satellites are a new, and growing, source of risk for radio observatories [e.g. 6,7]. Astronomers at the future site of the next-generation Square Kilometre Array in Australia have already detected intended and unintended emission from telecommunications satellites [8].

For this study, we examined the strength and distribution of terrestrial-based RF emission in and around Calgary, Alberta, and at the DRAO to characterize the RF emission distribution. While the RF environment has changed since we conducted this study in 2018, as a consequence of increased satellite activity and other forms of emission, we present these results as documentation of the past environment with the aim to redo the measurement. We aim to present a more general look at RFI as, often, observatories do not publicize their RFI measurements, or they are published in observatory-specific technical reports. In Section 2, we outline the experimental design, including the frequency bands and locations of the measurements, and the equipment used. In Section 3, we show the observational results (3.1-3.3) and discuss the strongest emission seen (3.4). In Section 4, we summarize the paper.

## EXPERIMENTAL DESIGN

At a combination of public and private, outdoor and indoor, and busy and quiet areas, we measured the radio spectrum in five frequency bands commonly used in communications and research. We took observations between March and June 2018 and determined the average spectra over two-minute intervals. Most measurements, aside from those at Mount Yamnuska, Nose Hill, and the apartment were taken during business hours. The apartment measurements were taken on a weekday evening, and the mountain and hill measurements on a Sunday during daytime. We aimed for these times to reflect the environment of everyday life.

### BANDS OF INTEREST

We chose two bands covering the cell phone ranges (824-960 MHz, 1710-2170 MHz) and an unlicensed ‘industrial, science, and medical devices’ (ISM) band (2400-2500 MHz; as in [9]). As radio astronomers, we also chose two radio astronomy bands (406-410 MHz, 1405-1435 MHz). These two frequency bands are the ones observed by the Synthesis Telescope at the DRAO [10].

### LOCATIONS

We chose several different locations around Calgary and Kananaskis, Alberta, (Figure 2) to investigate the variety of RF environments we could encounter in day-to-day life. At the University of Calgary, we chose four locations: MacEwan Hall, a public student hub (labeled as *Hub* in the figures); the Taylor Institute for Teaching and Learning, a moderately used, technologically advanced building (*Inst*); a semi-isolated basement lab (*Lab*); the outdoor quad in between the Administration and MacKimmie buildings, which is a well-frequented area with close proximity to four cell phone towers (*Quad*). We took three measurements elsewhere in the city: Nose Hill Park, a large, open, urban area (*Park*); an apartment downtown (*Apt*); a street in the Varsity neighbourhood (*Nbhd*). Additionally, we measured at three locations outside the city: a rural area off Highway 1A near Cochrane (*Rural*); on the front, prairie-facing side of Mount Yamnuska (*Mtn f*); on the backside of Mount Yamnuska (*Mtn b*). We also performed the same measurements at the DRAO (*DRAO*), which is a designated radio-quiet zone.

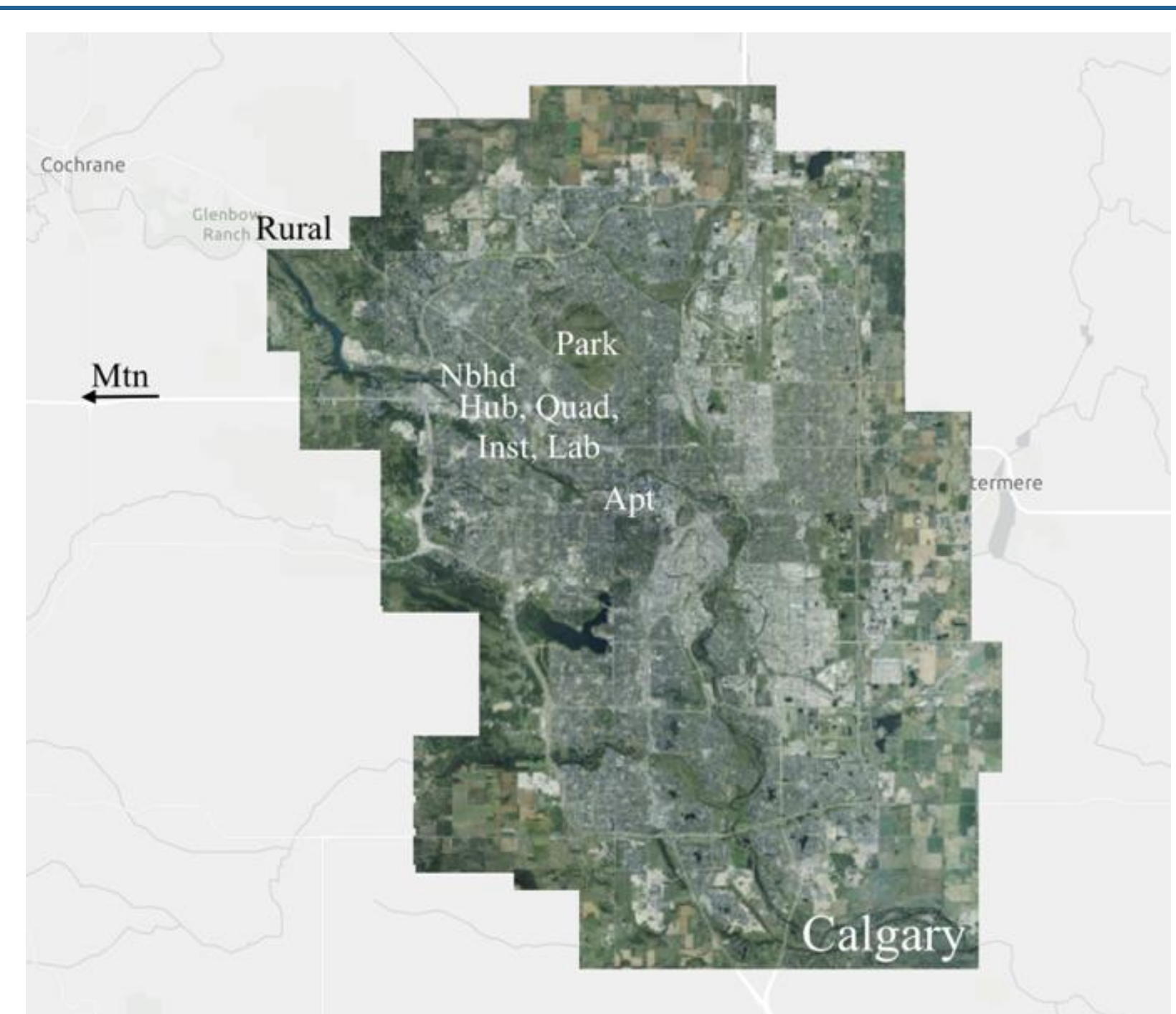

Figure 2. Locations in and around Calgary, Alberta, where we took ambient RF emission measurements. Each is labeled with its abbreviation used in the paper. Image credit City of Calgary, Esri Canada, Esri, TomTom, Garmin, SafeGraph, FAO, METI/NASA, USGS, EPA, NRCan, Parks Canada.

## EQUIPMENT AND PROCEDURE

We used a Keysight N9340B handheld RF spectrum analyzer [11] for measurements. With a range of 100 kHz to 3 GHz, narrow resolution bandwidths, and a preamplifier, this was an ideal choice for a portable, accurate device that could measure potentially weak signals.

We attached to the spectrum analyzer an Aaronia OmniLOG 90200 [12], which is a small radial isotropic broadband antenna. This antenna is ideal for our experiment as it has equal gain in all radial directions, allowing us to measure ambient RF signals without worrying about the source direction. Since the number of, and distance to, all RF-emitting devices was not determined, the ambient RF signals measured represent uncontrolled exposure situations.

We used the spectrum analyzer in ‘Spectrogram’ mode, which measures the average power continuously over time. Using this mode, we can determine the continuity and strength of signals, and determine the average and peak power one would encounter at each location. We adjusted the input attenuation depending on the signal strength of each location, increasing the attenuation for locations with strong emission. The attenuation level, if set to 10 or 20 dBm, results in an increase in the noise floor of the same value (as will be seen in the emission figures). We measured the noise for the instrument in each frequency band with a terminated input.

## RF EMISSION OBSERVATIONS

Below we present the average spectral distributions for our five bands of interest in the locations identified above. This section begins with the radio astronomy bands, followed by the cell phone bands, and then finally the unlicensed band. We present relative comparisons of the distributions. In addition to the average spectral distributions, we also recorded the maximum power intensities at the various locations in the different bands. We present the top five power intensities, and compare them to Health Canada exposure limits in the final subsection.

### RADIO ASTRONOMY BANDS (406-410 MHZ, 1405-1435 MHZ)

Figure 3 displays the average power of signals in the 406-410 MHz band. There were four specific signals prominent between 409-410 MHz. Notably, a 409.9 MHz signal was strong around campus (*Quad*, -85 dBm, *Hub*, -89 dBm), as well as in suburban Calgary (*Nbhd*, -79 dBm). A 409.1 MHz signal was prominent in the suburban and rural areas as well. The designated use for the band contains fixed and mobile radio use as well as radio astronomy continuum observations, where 'fixed' and 'mobile' are terms describing any radio-communication device between fixed or mobile stations, respectively [13]. Use in the 409-410 MHz sub-band contains public and private radio dispatch systems [14]. For example, signals seen on campus may come from campus security, grounds maintenance, or parking radios. The exact sources of the RF signals were not determined since these facilities do not wish to disclose their frequency of use and compromise their radios. Less public, or isolated, places (*Mtn b* and *DRAO*) showed no emission.

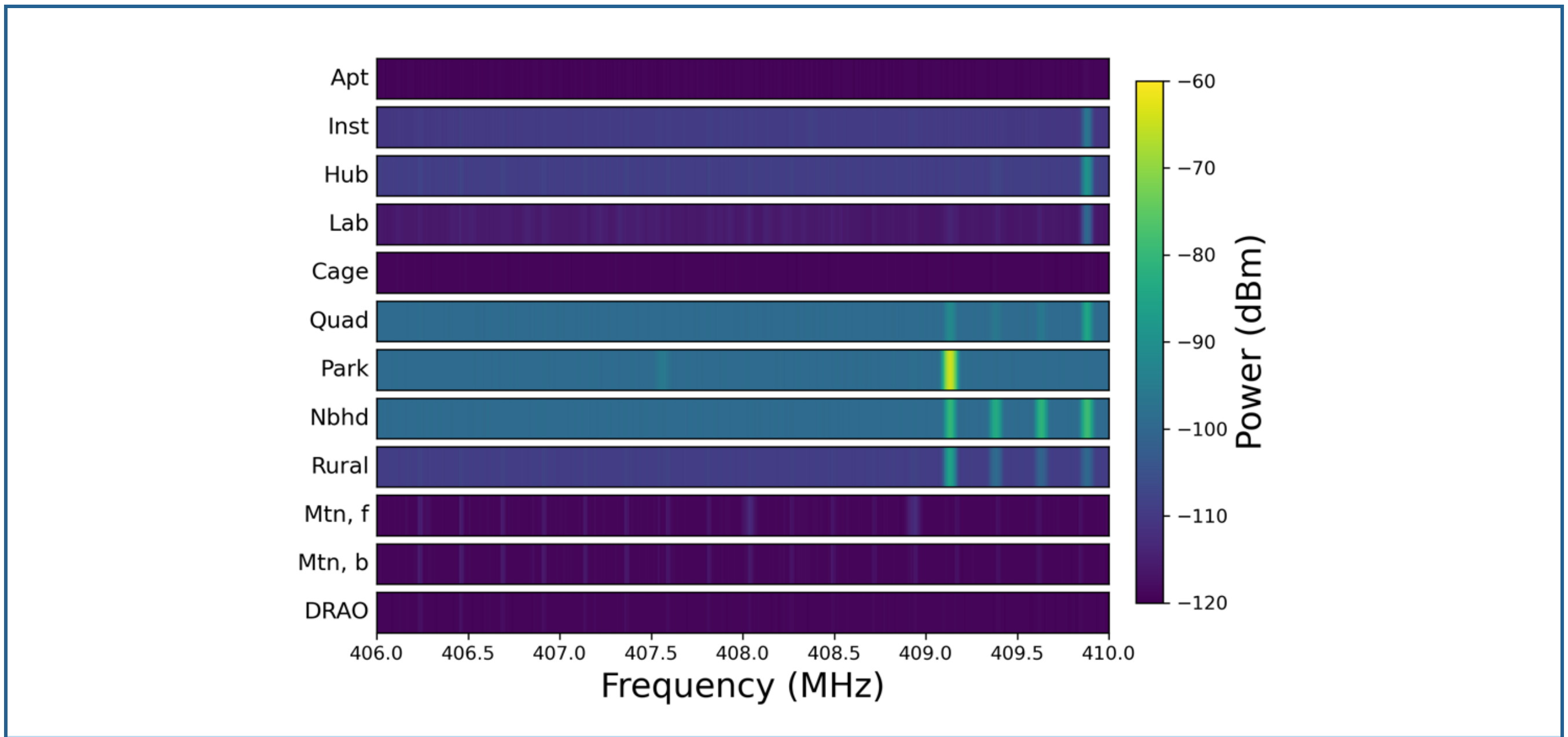

Figure 3. The power (dBm) measured at each location for the 406-410 MHz radio astronomy band. This band is also used for fixed and mobile use and tends to be used by public and private dispatch services. The measured noise levels at this frequency with the respective attention are: -119 dBm (*Apt, Lab, Cage, Mtn f, Mtn b, DRAO*); -109 dBm (*Inst, Hub, Rural*); -99 dBm (*Quad, Park, Nbhd*).

In the 1405-1435 MHz range, shown in Figure 4, there is a lack of signal across most bands. This is an expected result since, aside from radio astronomy use, the range 1400-1427 MHz has designation for passive earth exploration (via satellite) and space research, and thus no emissions are allowed at these frequencies [13]. This protected band is essential for mapping the small- and large-scale structure of the Galaxy since it corresponds to the 1420 MHz hyperfine transition line of neutral hydrogen, abundant in interstellar space [15]. There is a fixed and mobile devices band that begins at 1429 MHz, which may explain the faint emission seen in otherwise quiet areas (*Mtn f, Mtn b,* and *DRAO*) at 1433 MHz.

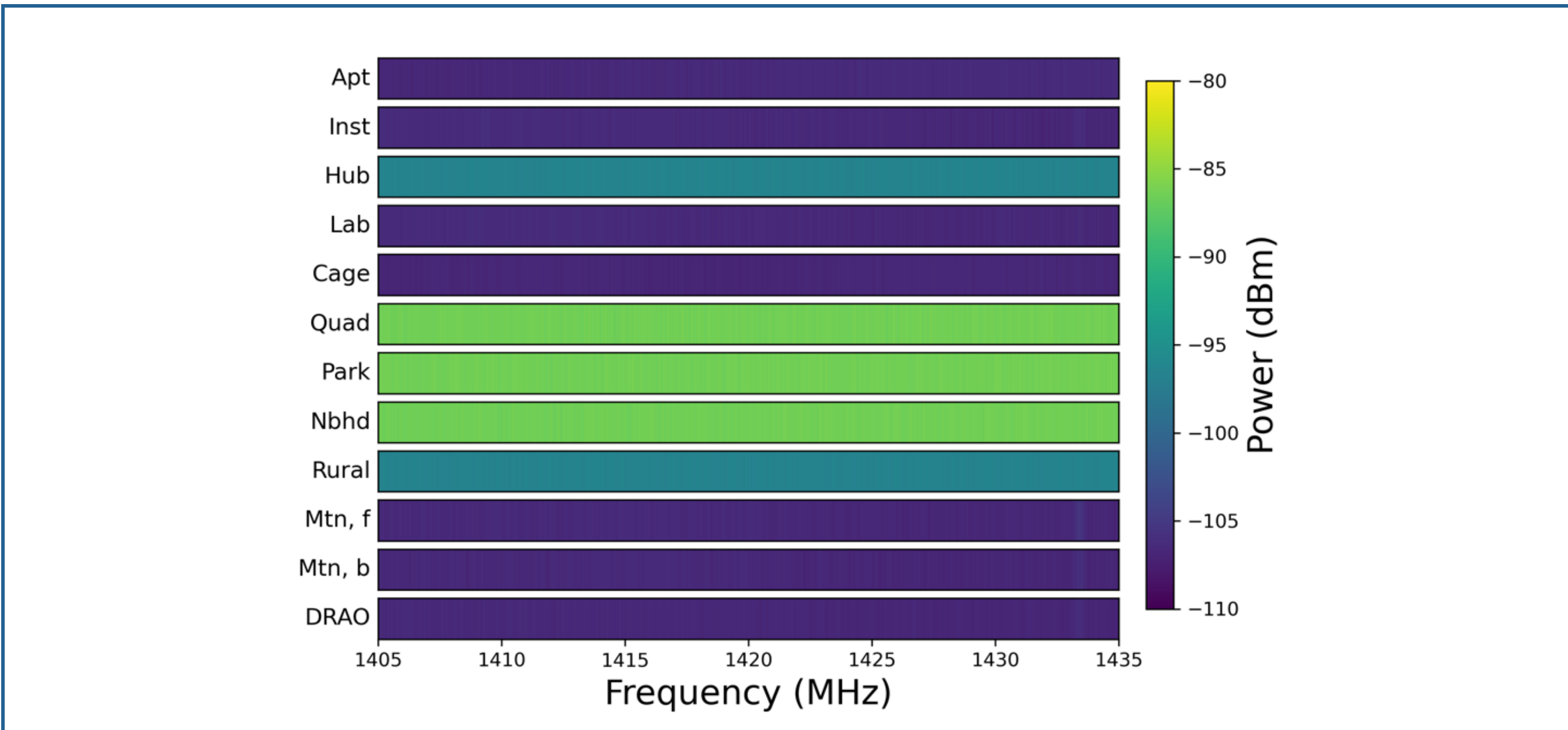

Figure 4. The power (dBm) measured at each location for the 1405-1435 MHz radio astronomy band. Note that the power scale is much lower in this figure compared to the others, and these measurements likely consist primarily of noise. The measured noise levels at this frequency with the respective attention are: -107 dBm (*Apt*, *Inst*, *Lab*, *Cage*, *Mtn f*, *Mtn b*, *DRAO*); -97 dBm (*Hub*, *Rural*); -86 dBm (*Quad*, *Park*, *Nbhd*).

## CELL PHONE BANDS (824-960 MHZ, 1710-2170 MHZ)

In the lower-frequency cell phone band (Figure 5), campus quad (*Quad*) shows the strongest signals at - 7 dBm, owing to the placement of four cell phone towers nearby. The Taylor Institute (*Inst*) shows emission across the entire band, unlike other locations, likely due to the prominence of licensed and amateur digital devices used to make it an interactive learning space. In remote locations, there is strong emission on the prairie exposed front side of Mount Yamnuska (*Mtn f*, -67 dBm), while the mountain sheltered landscapes have largely attenuated signals (*Mtn b*, -90 dBm, and *DRAO*, -94 dBm). The strong signals seen at many locations fall within the 869-894 MHz band allocated for cellular mobile use in the IMT system [16]. Cellular mobile systems also use the range 824-849 MHz, while trunked mobile radio systems, which control user traffic by automatically designating users to certain channels, radiolocation devices and amateur radio services fill the rest of this explored range [16].

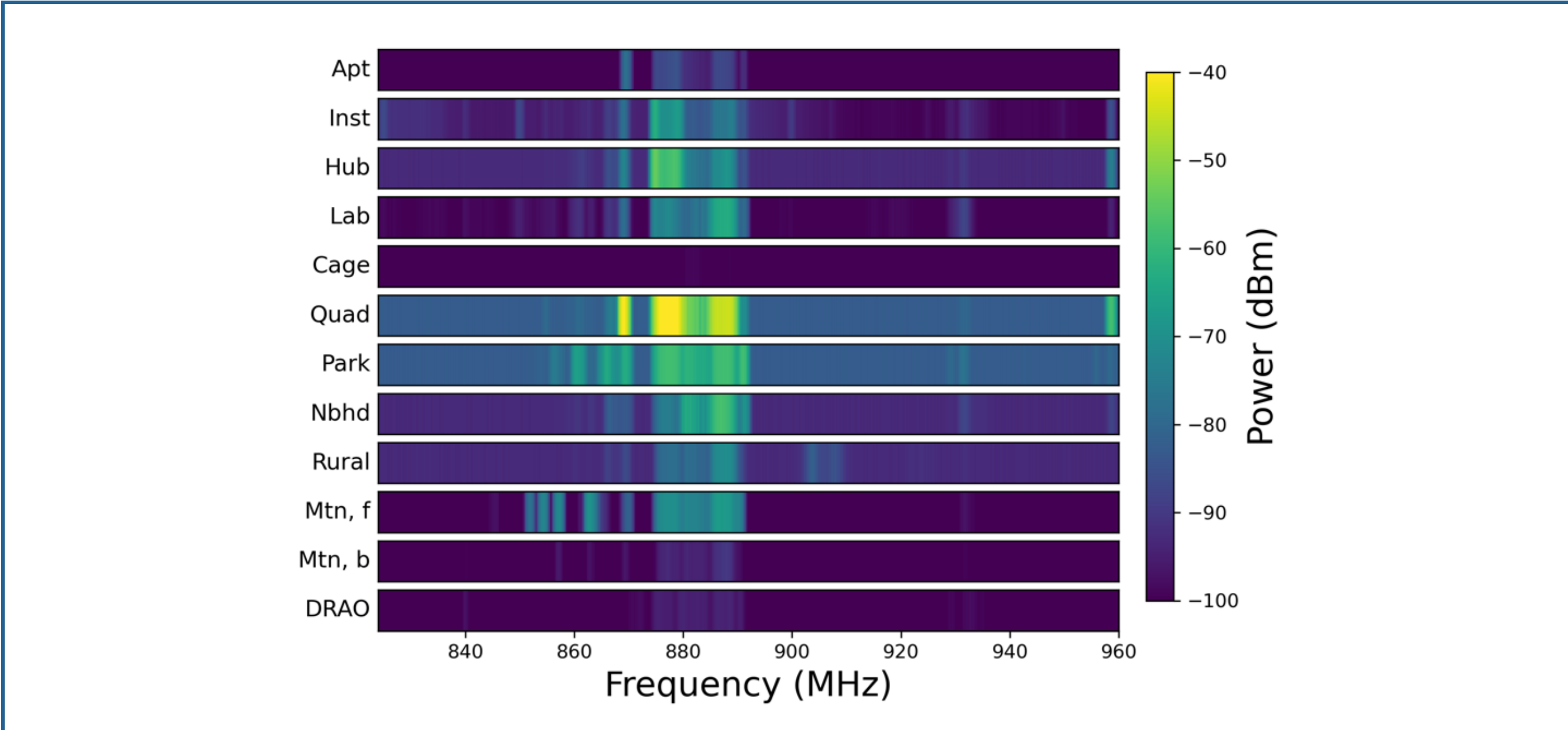

Figure 5. The power (dBm) measured at each location for the 824-960 MHz band. Mobile phones, amateur radio or radiolocation devices are some examples of other uses for this band. The measured noise levels at this frequency with the respective attention are: -103 dBm (*Apt*, *Lab*, *Cage*, *Mtn f*, *Mtn b*, *DRAO*); -93 dBm (*Inst*, *Hub*, *Nbhd*, *Rural*); -83 dBm (*Quad*, *Park*).

The higher-frequency cell phone band (Figure 6) also contains smaller sub-bands of fixed and mobile communication services. As illustrated in the figure, most devices around Calgary use the IMT-designated 1710-1755 MHz, 1850-2000 MHz, and 2110-2810 MHz bands [13, 17]. Similar to the lower cell phone range explored, the campus quad (Quad) displayed the strongest signals at -45 dBm. MacEwan Hall (*Hub*), Nose Hill Park (*Park*), and a neighbourhood street (*Nbhd*) also had quite strong peak strength between -45 and -60 dBm, while areas sheltered by the environment such as the apartment (*Apt*) and the front of Mount Yamnuska (*Mtn f*) showed weaker emission peaking near - 73 dBm.

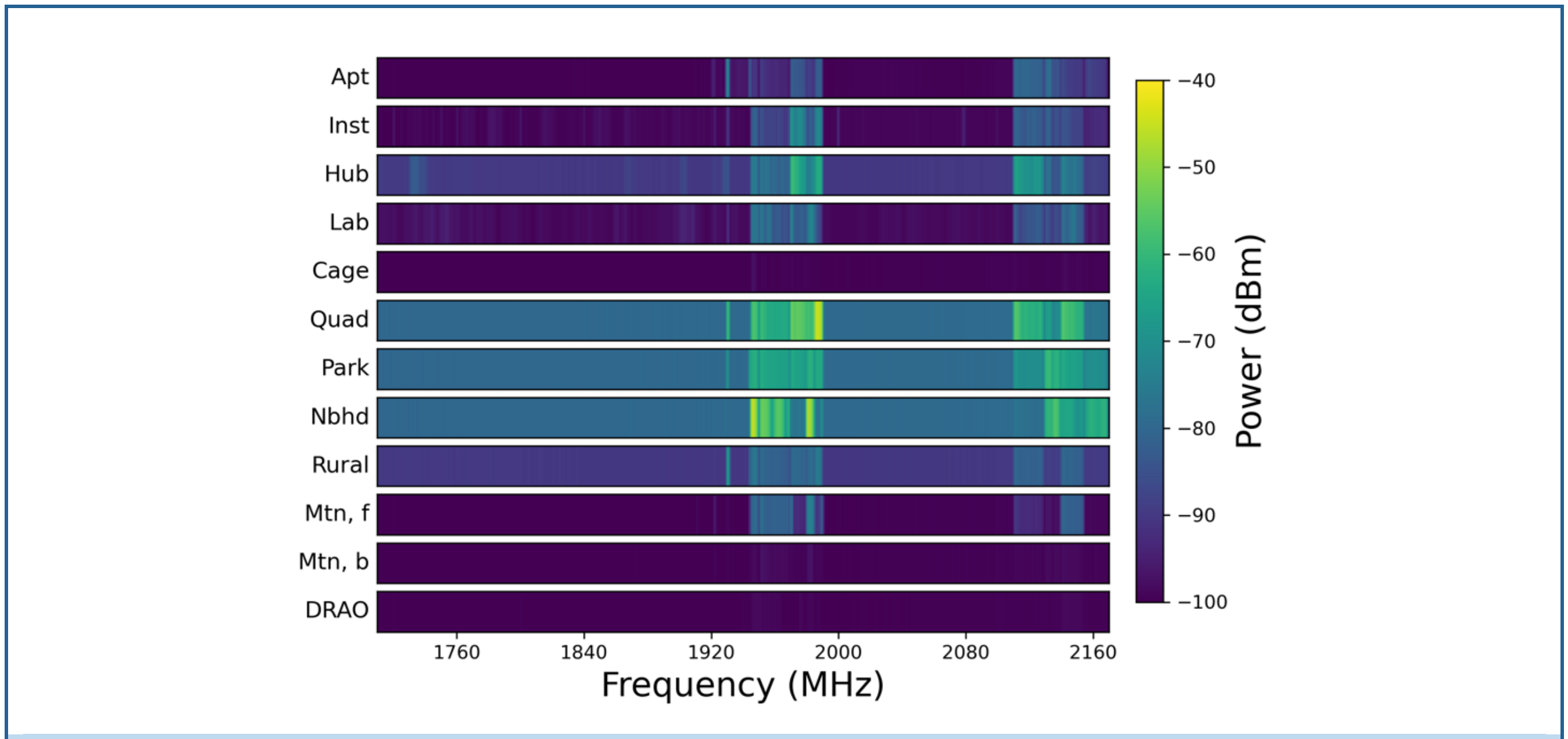

Figure 6. The power (dBm) measured at each location for the 1710-2170 MHz band. Emission in this band is mostly restricted to the specific frequencies that mobile phones use. The measured noise levels at this frequency with the respective attention are: -100 dBm (*Apt*, *Inst*, *Lab*, *Cage*, *Mtn f*, *Mtn b*, *DRAO*); -91 dBm (*Hub*, *Rural*); -80 dBm (*Quad*, *Park*, *Nbhd*).

In both bands, indoor spaces typically show decreased signal strength, likely due to attenuation from building materials such as concrete [18]. The architecture strongly influences the amount of RF emission seen. In outdoor spaces, the line of sight to emitting cell phone towers is a large factor in the strength of signal as they are consistently emitting at high power, and the power density decreases with the inverse square of distance [19, 20]. Consequently, the campus quad (Quad), a suburban neighbourhood (*Nbhd*), and the front side of Mount Yamnuska (*Mtn f*) are prime locations for strong, consistent signals. In certain outdoor (*Rural* and *Mtn f*) settings, the peak signals were over -20 dBm lower than the open area of the Quad. In other outdoor, more remote locations (*DRAO* and *Mtn b*), the signals peaked only 2-10 dBm above the noise. Here, natural barriers or large distances from emitting sources help provide large reductions in signal strength. Emissions are still faintly seen in the radio-quiet zone, demonstrating that protective measures are not strictly sufficient and astrophysical observations may easily be contaminated.

## UNLICENSED ISM DEVICES BAND (2400-2500 MHZ)

Figure 7 shows the spectra for ISM devices. Signals were only prominently seen in MacEwan Hall (*Hub*, -80 dBm) and the Taylor Institute (*Inst*, -88 dBm), and possibly seen in the apartment (*Apt*) and the lab (*Lab*). Many devices that use this band emit intermittently. The measured spectra can show variation and are dependent on the surroundings at the exact time of measurement. However, Kwan and Fapojuwo [9] note that despite intermittent signals, they find overall radio emissions are not significantly different between peak and off-peak hours in the cell phone and ISM devices bands. The ISM Devices band is also used for fixed, mobile and radiolocation devices [13]. However, its unlicensed nature means that Wi-Fi, Bluetooth, cordless phones, microwaves, and many other non-cellular devices exploit this band. This is a consequence of the limited spectrum and cost of the licensed spectrum which has driven companies towards network services, such as Wi-Fi, in this band and others [21].

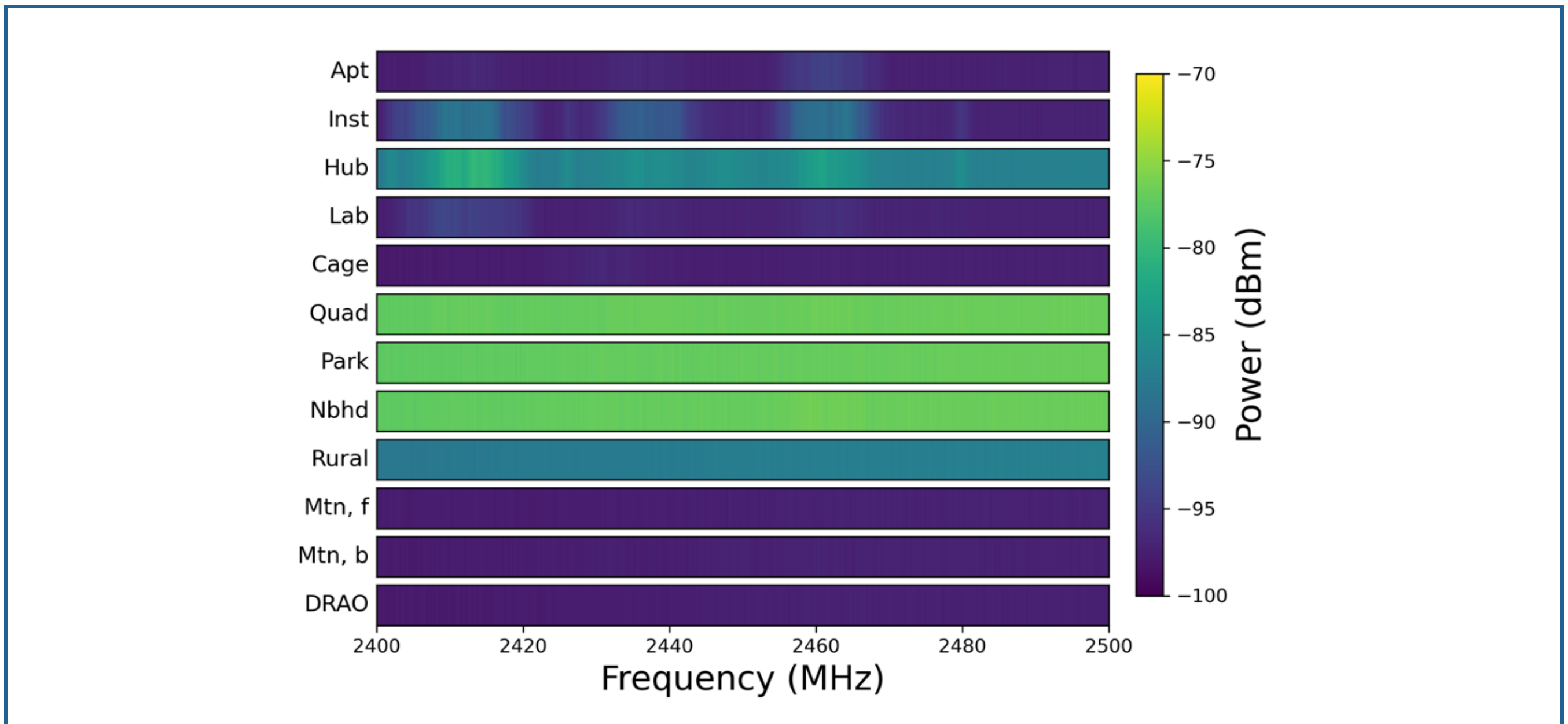

Figure 7. The power (dBm) measured at each location for the 2400-2500 MHz band. This contains industrial, science, and medical devices. The measured noise levels at this frequency with the respective attention are: -97 dBm (*Apt*, *Inst*, *Lab*, *Cage*, *Mtn f*, *Mtn b*, *DRAO*); -87 dBm (*Hub*, *Rural*); - 7 dBm (*Quad*, *Park*, *Nbhd*).

## POWER INTENSITY

In Table 1 we present the power intensities for the five strongest signals we observed, all of which are in the cell phone bands, and compare these powers to the exposure limits published by Health Canada. To calculate the power intensity, $S$ [W m$^{-2}$], we use

$$S = \frac{10^{P/10}}{1000} \frac{4\pi}{\lambda^2 G} \ ,$$

where $P$ is the measured power [dBm], $\lambda$ is the wavelength of interest [m], and $G$ is the gain of the antenna. The gain of the antenna is found by calculating

$$G = 10^{\frac{g}{10}} \ ,$$

where $g$ is the gain in dBi (decibel in relation to an isotropic antenna), which is provided by the manufacturer for all frequencies in the antenna's range. This calculation is relating the received power from the effective aperture of the antenna and assumes the antenna is receiving the maximum power from the incoming radiation.

### TABLE 1

Power intensity [in W $m^{-2}$] of the five strongest signals we measured. The signals were averaged across a 2-minute observation. Health Canada exposure limits are shown for the relevant frequency.

| Location | Frequency (MHz) | Avg. Power (dBm) | Std. Dev. (dBm) | Avg. Power ($10^{-6}$ W $m^{-2}$) | Health Canada Limit (W $m^{-2}$) |
|---|---|---|---|---|---|
| Quad | 876 | -37.76 | 3.16 | 17.3 | 2.69 |
| Quad | 1986 | -45.02 | 2.95 | 12.5 | 4.70 |
| Nbhd | 1945 | -45.72 | 1.51 | 9.4 | 4.63 |
| Hub | 1971 | -59.33 | 5.09 | 0.4 | 4.67 |
| Inst | 1970 | -59.54 | 3.19 | 0.5 | 4.67 |

Exposure limits are determined by Health Canada from numerous studies investigating the effects of RF exposure on the human body [1]. The exposure limit is determined as the lowest strength of field for which there has been an identified health concern, with a margin for safety. The calculated limits for a 6-minute exposure, in power density $S_{RL}$ [W $m^{-2}$], for the 300 MHz to 6 GHz range can be expressed by

$$S_{RL} = 0.02619 f^{0.6834} \ ,$$

where $f$ is the frequency of interest, in units of MHz [1].

The strongest power density we measured is more than a million times lower than the exposure limit. Similarly, Kwan and Fapojuwo [9] also found the strongest consistent signal to be in an open area near a cell phone tower with a power density of $4 \times 10^{-5}$ W $m^{-2}$. While this is comparable to our strongest signal (Table 1), they are both still much smaller than the recommended safe limit.

For comparison, one of the strongest extragalactic radio signals, the supernova remnant Cassiopeia A, was $2.7 \times 10^{-23}$ W $m^{-2}$ $Hz^{-1}$ at 1 GHz in 1980 and has been decreasing in strength over time [22]. Many other radio sources are on the order of $10^{-25}$ to $10^{-24}$ W $m^{-2}$ $Hz^{-1}$ [23]. The units include $Hz^{-1}$, indicating the detector bandwidth must be taken into consideration. For example, if a detector has a 1 GHz bandwidth, the power density would then be around $10^{-14}$ W $m^{-2}$. Clearly, astronomical signals are much weaker than anthropogenic signals, justifying the protective measures and radio-quiet zones in place at radio observatories.

## CONCLUSIONS

The radio window of our atmosphere is partitioned and allocated for use, including cellphone communication, satellite television, and research in space and astronomy. As radio astronomers, we were curious about the RF intensity present in our daily lives, and how RF intensities generated by industries compared to the signals we study from outer space.

We presented the average spectral distribution and maximum power intensities for five RF bands covering well-used portions of the radio spectrum in several different environments. The spectral distributions found were variable and environment dependent. What we found particularly interesting was that many cell phone bands had signals that were stronger in outdoor, public locations than in indoor environments. By contrast, signals in the unlicensed 'industrial, science and medical devices' band we investigated were stronger in indoor public places rather than outdoors. Finally, in the radio astronomy bands we explored, it was clear that radio-quiet zones at astronomical observatories are essential in efforts to reduce radio frequency interference, especially since portions of the radio spectrum allocated for radio astronomy are also used by various private services.

At radio observatories, there is continuous monitoring and mitigation of RFI. Current efforts are being directed towards the risk of satellite-produced interference — technology that is changing the feasibility of ground-based astronomy. How, then, are these satellites changing the radio exposure of our everyday environments?

## ACKNOWLEDGEMENTS

PF and JB would like to thank Dr. Anna Ordog, Dr. Mehrnoosh Tahani, and Dr. Cameron van Eck for their help in the analysis and interpretation of the results, and Dr. Tom Landecker, Dr. Alex Hill, and Dr. Andrew Gray for discussing these results in connection to DRAO. PF would like to thank Becky Booth and Laura Lewis for their help in hiking and collecting data.